\documentclass[conference]{IEEEtran}
\usepackage{cite}
\usepackage{amsmath,amssymb,amsfonts}
\usepackage{pifont}
\newcommand{\xmark}{\ding{55}}
\usepackage{algorithmic}
\usepackage{graphicx}
\usepackage{textcomp}
\usepackage{xcolor}
\usepackage{wrapfig}
\usepackage{comment}
\usepackage{amsthm}
\usepackage{longtable}
\usepackage{ upgreek }
\usepackage{amsmath}
\usepackage{graphicx} 
\usepackage{ gensymb }
\usepackage{ dsfont }
\usepackage{tabularx,booktabs}
\usepackage[acronym,nomain,nonumberlist]{glossaries}
\usepackage{glossary-tree}
\usepackage{tabularx}
\usepackage{float}
\newcolumntype{L}{>{\raggedright\arraybackslash}X}

\usepackage{cleveref}
\usepackage{subcaption}
\usepackage{makecell}
\usepackage{multirow}
\usepackage{subcaption}
\usepackage{fancyhdr}

\usepackage[ruled,norelsize]{algorithm2e}
\makeatletter
\newcommand{\removelatexerror}{\let\@latex@error\@gobble}
\makeatother
\SetKw{And}{\hspace{\algoskipindent}\itshape and\;}
\SetKwBlock{Condition}{}{}
\SetKw{Or}{\hspace{\algoskipindent}\itshape or\;}

\usepackage[top=0.75in, bottom=1in, left=0.625in, right=0.625in]{geometry}
\makeglossaries
\newacronym{mmtc}{mMTC}{massive machine-type com\-mu\-ni\-ca\-tions}
\newacronym{urllc}{URLLC}{ultra-reliable low latency communications}
\newacronym{embb}{eMBB}{enhanced mobile broadband}
\newacronym{ntn}{NTN}{non-terrestrial network}
\newacronym{tn}{TN}{terrestrial network}
\newacronym{hrcs}{\hbox{HRC-s}}{high reliability communications via satellite}
\newacronym{embbs}{\hbox{eMBB-s}}{enhanced mobile broadband via satellite}
\newacronym{mmtcs}{\hbox{mMTC-s}}{massive machine-type communications via satellite}
\newacronym{3gpp}{3GPP}{3rd Generation Partnership Project}
\newacronym{leo}{LEO}{low Earth orbit}
\newacronym{meo}{MEO}{medium Earth orbit}
\newacronym{geo}{GEO}{geosynchronous orbit}
\newacronym{haps}{HAPS}{high-altitude platform station}
\newacronym{nr}{NR}{New Radio}
\newacronym{vsat}{VSAT}{very small aperture terminal}
\newacronym{tr}{TR}{technical report}
\newacronym{ts}{TS}{technical specification}
\newacronym{frf}{FRF}{frequency reuse factor}
\newacronym{rx}{Rx}{receive}
\newacronym{tx}{Tx}{transmit}
\newacronym{sinr}{SINR}{signal-to-interference-plus-noise ratio}
\newacronym{mrc}{MRC}{maximum ratio combining}
\newacronym{ue}{UE}{user equipment}
\newacronym{rhcp}{RHCP}{right-hand circular polarization}
\newacronym{lhcp}{LHCP}{left-hand circular polarization}
\newacronym{ns3}{ns-3}{network simulator 3}
\newacronym{rng}{RNG}{random number generator}
\newacronym{dl}{DL}{downlink}
\newacronym{ul}{UL}{uplink}
\newacronym{imt2020}{\hbox{IMT-2020}}{International Mobile Telecommunications 2020}

\newacronym{}{}{}

\def\BibTeX{{\rm B\kern-.05em{\sc i\kern-.025em b}\kern-.08em
    T\kern-.1667em\lower.7ex\hbox{E}\kern-.125emX}}

\renewcommand\IEEEkeywordsname{Keywords}

\makeatletter
\newcommand{\linebreakand}{%
  \end{@IEEEauthorhalign}
  \hfill\mbox{}\par
  \mbox{}\hfill\begin{@IEEEauthorhalign}
}
\makeatother

\IEEEoverridecommandlockouts 

\fancypagestyle{FirstPage}{
\lfoot{This work has been submitted to the IEEE for possible publication. Copyright may be transferred without notice, after which this version may no longer be accessible.} 
}

\begin{document}
\bstctlcite{IEEEexample:BSTcontrol}
\bstctlcite{bstctl:etal, bstctl:nodash, bstctl:simpurl}

\newtheorem{thm}{Theorem} 
\theoremstyle{definition}
\newtheorem{remark}[thm]{Remark}
\newtheorem{defn}[thm]{Definition}
\theoremstyle{plain}
\newtheorem{thr}[thm]{Theorem}
\newtheorem{prop}[thm]{Proposition}
\newtheorem{kor}[thm]{Corollary}

\title{Can 3GPP New Radio Non-Terrestrial Networks Meet the IMT-2020 Requirements for Satellite Radio Interface Technology?}

\author{
\thanks{This work has been funded by the European Space Agency project HELENA (Highly skillEd sateLlite community mEmbers to drive 3GPP Non-Terrestrial Network stAndardization) - Support to Standardisation of Satellite 5G Component under program ARTES 4.0 Core Competitiveness Generic Programme Line – Future Preparation. The views expressed are those of the authors and can in no way be taken to reflect the official opinion of the European Space Agency.}

\IEEEauthorblockN{Mikko Majamaa\IEEEauthorrefmark{1}\IEEEauthorrefmark{2}, Lauri Sormunen\IEEEauthorrefmark{1}, Verneri Rönty\IEEEauthorrefmark{1}, Henrik Martikainen\IEEEauthorrefmark{1}, Jani Puttonen\IEEEauthorrefmark{1},\\ and Timo Hämäläinen\IEEEauthorrefmark{2}}

\IEEEauthorblockA{
\IEEEauthorrefmark{1}\textit{Magister Solutions, Jyv\"{a}skyl\"{a}, Finland} \\
email: \{mikko.majamaa, lauri.sormunen, verneri.ronty, henrik.martikainen, jani.puttonen\}@magister.fi
}

\IEEEauthorblockA{
\IEEEauthorrefmark{2}\textit{Faculty of Information Technology, University of Jyv\"{a}skyl\"{a}, Finland} \\
email: timo.t.hamalainen@jyu.fi
}

}

\maketitle

\begin{abstract}

The International Telecommunication Union defined the requirements for 5G in the \gls{imt2020} standard in 2017. Since then, advances in technology and standardization have made the ubiquitous deployment of 5G via satellite a practical possibility, for example, in locations where \glspl{tn} are not available. However, it may be difficult for satellite networks to achieve the same performance as \glspl{tn}. To address this, the \gls{imt2020} requirements for satellite radio interface technology have recently been established. In this paper, these requirements are evaluated through system simulations for the 3rd Generation Partnership Project New Radio non-terrestrial networks with a low Earth orbit satellite. The focus is on the throughput, area traffic capacity, and spectral efficiency requirements. It is observed that the downlink (DL) requirements can be met for user equipment with 2~receive antenna elements. The results also reveal that frequency reuse factor~1 (FRF1) may outperform FRF3 in DL with a dual-antenna setup, which is a surprising finding since FRF3 is typically considered to outperform FRF1 due to better interference reduction. For \gls{ul}, 1~transmit antenna is sufficient to meet the requirements by a relatively large margin – a promising result given that \gls{ul} is generally more demanding.

\end{abstract}


\section{Introduction}
\label{sec:introduction}

\thispagestyle{FirstPage}

\glsresetall

The \gls{imt2020} standard \cite{ITU-RM2083}, published by the International Telecommunication Union (ITU), defines the requirements for 5G networks, devices, and services. The standard covers \gls{urllc}, \gls{embb}, and \gls{mmtc} usage scenarios.

\begin{figure}[htb!]
    \centering
    \includegraphics[width=.89\linewidth]{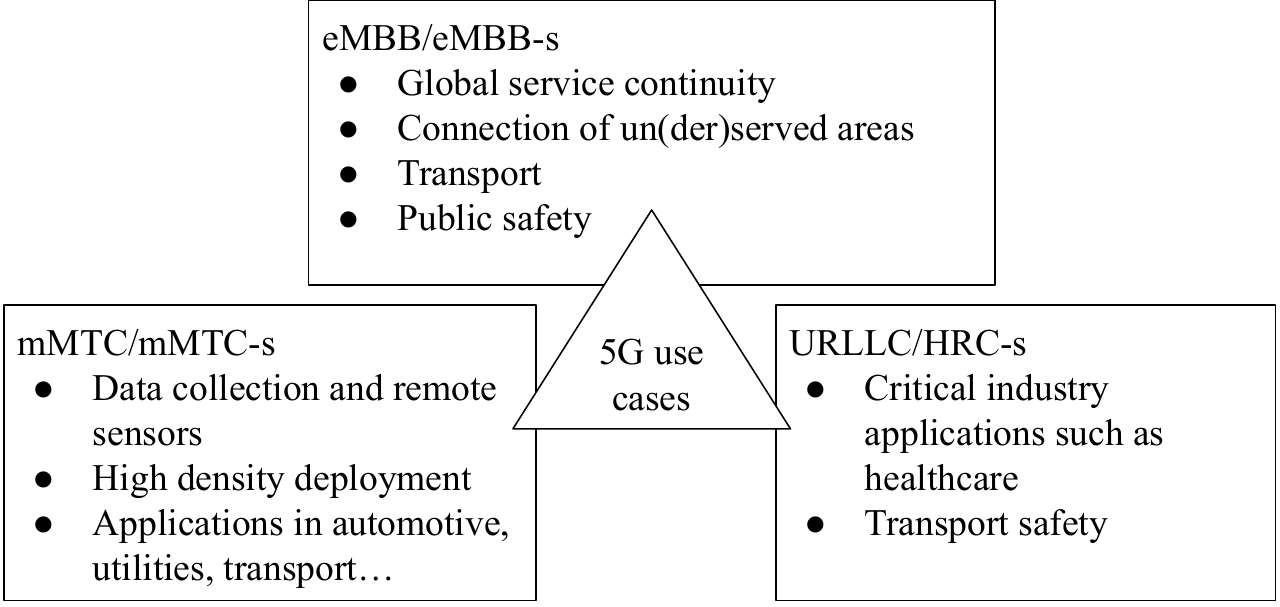}
    \caption{5G use cases. Adapted from \cite{ITU-RM2514}.}
    \label{fig:5g_use_cases}
\end{figure}

Because \glspl{tn} are not available everywhere, satellite communications can be used to provide connectivity to areas where it would otherwise be impossible, costly, or hazardous. In addition, a satellite component can be used for load balancing or in the event of crises when \glspl{tn} are out of service. However, as identified by ITU, the requirements set for 5G will need to be adapted for satellite radio interface technology, taking into account characteristics such as higher latency and possibly lower bandwidth. Therefore, ITU has defined requirements for satellite radio interfaces of \gls{imt2020} in \hbox{ITU-R M.2514-0} \cite{ITU-RM2514}. These requirements include \gls{hrcs}, \gls{embbs}, and \gls{mmtcs} usage scenarios. Requirements for \gls{hrcs}, \gls{embbs}, and \gls{mmtcs} \cite[\hbox{Table 8.2.6.3}]{ITU-RM2514} are less stringent than their terrestrial counterparts \cite[Table~II]{8985528}. The 5G use cases for terrestrial/satellite components are illustrated in Fig.~\ref{fig:5g_use_cases}.

{}



The \gls{3gpp} \cite{3gpp} is an umbrella term for several standardization organizations that provide specifications for mobile communications. The work in 3GPP is organized in releases with cycles of about 18 months. Releases 15 through 17 focused on 5G, while the ongoing Release~18 (\hbox{Rel-18}) marks the beginning of 5G-Advanced (\hbox{5G-A}). Rel-17 was historic because it included satellite communications, or \glspl{ntn} in 3GPP nomenclature, in the 3GPP specifications for the first time, although, \hbox{Rel-15} and \hbox{Rel-16} included necessary preparatory work in the form of \glspl{tr} and study items \cite{38811, 38821}. The significance of the standardization efforts lies in cooperation – it is expected that \glspl{tn} and \glspl{ntn} will become a unified network \cite{9914764}. From the user’s point of view, this means an indistinguishable network, that is, the user could be connected to a network without noticing whether it is via a terrestrial or a satellite access point.

The \gls{ntn} work in 3GPP covers \gls{leo}, \gls{meo}, and \gls{geo} satellite communications but also airborne vehicles such as \glspl{haps}. \hbox{Rel-17} included support for the basic funtionality of \gls{nr}, the air interface of 5G, through satellites in \hbox{S- and L-bands} serving handheld terminals. \hbox{Rel-18} extends \gls{nr} capabilities through \glspl{ntn}, for example, by considering higher frequencies, mobility enhancements, and support for \glspl{vsat} \cite{rel18overview}.


In the past, several works have considered the performance of \glspl{ntn}. In \cite{9347998}, the authors evaluate the capacity and throughput of \gls{ntn} with \gls{leo} satellites. The throughput performance of \gls{ntn} with \gls{geo} satellite is considered in \cite{9289577}. The handover performance in \gls{leo} \gls{ntn} is considered in \cite{9906486}. In \cite{9685392}, the authors evaluate the performance of \gls{mmtc} over a \gls{leo} satellite. Although evaluations of \glspl{ntn} have been done in the past, to the best of the authors’ knowledge no published work comprehensively evaluates 3GPP NR NTNs against the \gls{imt2020} requirements for satellite radio interface technology. It is therefore important to study whether the requirements can be met by 3GPP NR NTNs, which is critical to ensure the deliverability of anticipated services and to provide strategic adaptation in cases where these networks fail to meet the requirements.

This paper aims to fill the identified research gap by providing an evaluation of NR over NTN in relation to the ITU’s requirements, focusing on the throughput, spectral efficiency, and area traffic capacity requirements in non-mobile \hbox{eMBB-s} scenarios. For the other requirements, the reader is referred to the results of the other partners in the research project, which indeed suggest that 3GPP NR NTNs can meet the ITU's requirements \cite{3GPPR1-2310940}.

The remainder of the paper is organized as follows. In the next section, the \gls{imt2020} requirements for satellite radio interface technology are outlined. In Section~\ref{sec:evaluations}, the requirements are evaluated through system-level simulations. Section~\ref{sec:conclusion} concludes the paper.

\section{IMT-2020 Requirements for Satellite Radio Interface Technology}
\label{sec:reqs}

Table~\ref{table:reqs}. outlines the minimum technical requirements for the 5G satellite radio interface set as defined by ITU. It also lists the evaluation methods, which include analysis, simulation, and inspection, depending on the requirement. Where applicable, the usage scenario, test environment, and link direction are listed. The last column shows which of the requirements are considered in this paper. The focus is on non-mobile eMBB-s throughput, spectral efficiency, and area traffic capacity statistics.

\begin{table*}[hbt!]
\begin{center}
\caption{Minimum technical requirements of the 5G satellite radio interface set by ITU.}
\label{table:reqs}
\begin{tabular}{|l|l|lll|l|l|}
\hline
\multirow{2}{*}{\textbf{\begin{tabular}[c]{@{}l@{}}Minimum technical \\ requirement item\end{tabular}}} & \multirow{2}{*}{\textbf{\begin{tabular}[c]{@{}l@{}}High-level assessment\\ method\end{tabular}}} & \multicolumn{3}{c|}{\textbf{Category}} & \multirow{2}{*}{\textbf{\begin{tabular}[c]{@{}l@{}}Required\\ value\end{tabular}}} & \multirow{2}{*}{\textbf{\begin{tabular}[c]{@{}l@{}}Included\\ in this\\ study\end{tabular}}} \\ \cline{3-5}
 &  & \multicolumn{1}{l|}{\textbf{\begin{tabular}[c]{@{}l@{}}Usage\\ scenario\end{tabular}}} & \multicolumn{1}{l|}{\textbf{\begin{tabular}[c]{@{}l@{}}Test\\ environment\end{tabular}}} & \textbf{\begin{tabular}[c]{@{}l@{}}\gls{dl} or\\ \gls{ul}\end{tabular}} &  &  \\ \hline
\multirow{2}{*}{Peak data   rate} & \multirow{2}{*}{Analytical} & \multicolumn{1}{l|}{\multirow{2}{*}{eMBB-s}} & \multicolumn{1}{l|}{\multirow{2}{*}{N/A}} & \gls{ul} & 2 Mbit/s &  \\ \cline{5-7} 
 &  & \multicolumn{1}{l|}{} & \multicolumn{1}{l|}{} & \gls{dl} & 70 Mbit/s &  \\ \hline
\multirow{2}{*}{\begin{tabular}[c]{@{}l@{}}Peak spectral  \\ efficiency\end{tabular}} & \multirow{2}{*}{Analytical} & \multicolumn{1}{l|}{\multirow{2}{*}{eMBB-s}} & \multicolumn{1}{l|}{\multirow{2}{*}{N/A}} & \gls{ul} & 1.5 bit/s/Hz &  \\ \cline{5-7} 
 &  & \multicolumn{1}{l|}{} & \multicolumn{1}{l|}{} & \gls{dl} & 3 bit/s/Hz &  \\ \hline
\multirow{2}{*}{\begin{tabular}[c]{@{}l@{}}User experienced\\ data rate\end{tabular}} & \multirow{2}{*}{\begin{tabular}[c]{@{}l@{}}Simulation and analytical\end{tabular}} & \multicolumn{1}{l|}{\multirow{2}{*}{eMBB-s}} & \multicolumn{1}{l|}{\multirow{2}{*}{Rural}} & \gls{ul} & 100 kbit/s & \checkmark \\ \cline{5-7} 
 &  & \multicolumn{1}{l|}{} & \multicolumn{1}{l|}{} & \gls{dl} & 1 Mbit/s & \checkmark \\ \hline
\multirow{2}{*}{\begin{tabular}[c]{@{}l@{}}5th percentile user \\ spectral efficiency\end{tabular}} & \multirow{2}{*}{Simulation} & \multicolumn{1}{l|}{\multirow{2}{*}{eMBB-s}} & \multicolumn{1}{l|}{\multirow{2}{*}{Rural}} & \gls{ul} & 0.003 bit/s/Hz & \checkmark \\ \cline{5-7} 
 &  & \multicolumn{1}{l|}{} & \multicolumn{1}{l|}{} & \gls{dl} & 0.03 bit/s/Hz & \checkmark \\ \hline
\multirow{2}{*}{\begin{tabular}[c]{@{}l@{}}Average spectral\\ efficiency\end{tabular}} & \multirow{2}{*}{Simulation} & \multicolumn{1}{l|}{\multirow{2}{*}{eMBB-s}} & \multicolumn{1}{l|}{\multirow{2}{*}{Rural}} & \gls{ul} & 0.1 bit/s/Hz & \checkmark \\ \cline{5-7} 
 &  & \multicolumn{1}{l|}{} & \multicolumn{1}{l|}{} & \gls{dl} & 0.5 bit/s/Hz & \checkmark \\ \hline
\multirow{2}{*}{Area traffic capacity} & \multirow{2}{*}{\begin{tabular}[c]{@{}l@{}}Simulation and analytical\end{tabular}} & \multicolumn{1}{l|}{\multirow{2}{*}{eMBB-s}} & \multicolumn{1}{l|}{\multirow{2}{*}{Rural}} & \gls{ul} & 1.5 kbit/s/km² & \checkmark \\ \cline{5-7} 
 &  & \multicolumn{1}{l|}{} & \multicolumn{1}{l|}{} & \gls{dl} & 8 kbit/s/km² & \checkmark \\ \hline
User Plane   latency & \begin{tabular}[c]{@{}l@{}}Analytical and inspection\end{tabular} & \multicolumn{1}{l|}{eMBB-s} & \multicolumn{1}{l|}{N/A} & N/A & 10 ms &  \\ \hline
Control Plane latency & \begin{tabular}[c]{@{}l@{}}Analytical and inspection\end{tabular} & \multicolumn{1}{l|}{eMBB-s} & \multicolumn{1}{l|}{N/A} & N/A & 40 ms &  \\ \hline
Connection density & Simulation & \multicolumn{1}{l|}{mMTC-s} & \multicolumn{1}{l|}{Rural} & N/A & 500 devices/km² &  \\ \hline
Energy efficiency & Inspection & \multicolumn{1}{l|}{eMBB-s} & \multicolumn{1}{l|}{N/A} & N/A & \begin{tabular}[c]{@{}l@{}}High sleep ratio and\\ long sleep duration\end{tabular} &  \\ \hline
Reliability & Simulation & \multicolumn{1}{l|}{HRC-s} & \multicolumn{1}{l|}{Rural} & N/A & 0.999 &  \\ \hline
Mobility – UE speed & Simulation & \multicolumn{1}{l|}{eMBB-s} & \multicolumn{1}{l|}{Rural} & N/A & 250 km/h &  \\ \hline
\begin{tabular}[c]{@{}l@{}}Mobility – Traffic channel\\ link data rate\end{tabular} & Simulation & \multicolumn{1}{l|}{eMBB-s} & \multicolumn{1}{l|}{Rural} & N/A & 0.005 bit/s/Hz &  \\ \hline
Mobility interruption time & Analytical & \multicolumn{1}{l|}{eMBB-s} & \multicolumn{1}{l|}{N/A} & N/A & 50 ms &  \\ \hline
Bandwidth & Inspection & \multicolumn{1}{l|}{N/A} & \multicolumn{1}{l|}{N/A} & N/A & \begin{tabular}[c]{@{}l@{}}At least up to and \\ including 30 MHz\end{tabular} &  \\ \hline
\end{tabular}
\end{center}
\end{table*}

\section{Evaluation of the Requirements}
\label{sec:evaluations}
\subsection{System Model}
\label{sec:systemmodel}

This section details the system model for evaluating the \hbox{eMBB-s} requirements in non-mobile cases through system-level simulations. The evaluation parameters are given in Table~\ref{table:params}. It should be noted that these parameters can be found in \hbox{ITU-R M.2514-0} but also that these parameters are aligned with the calibration parameters found in TR~38.821.

\begin{figure}[hbt!]
    \centering
    \includegraphics[width=.8\linewidth]{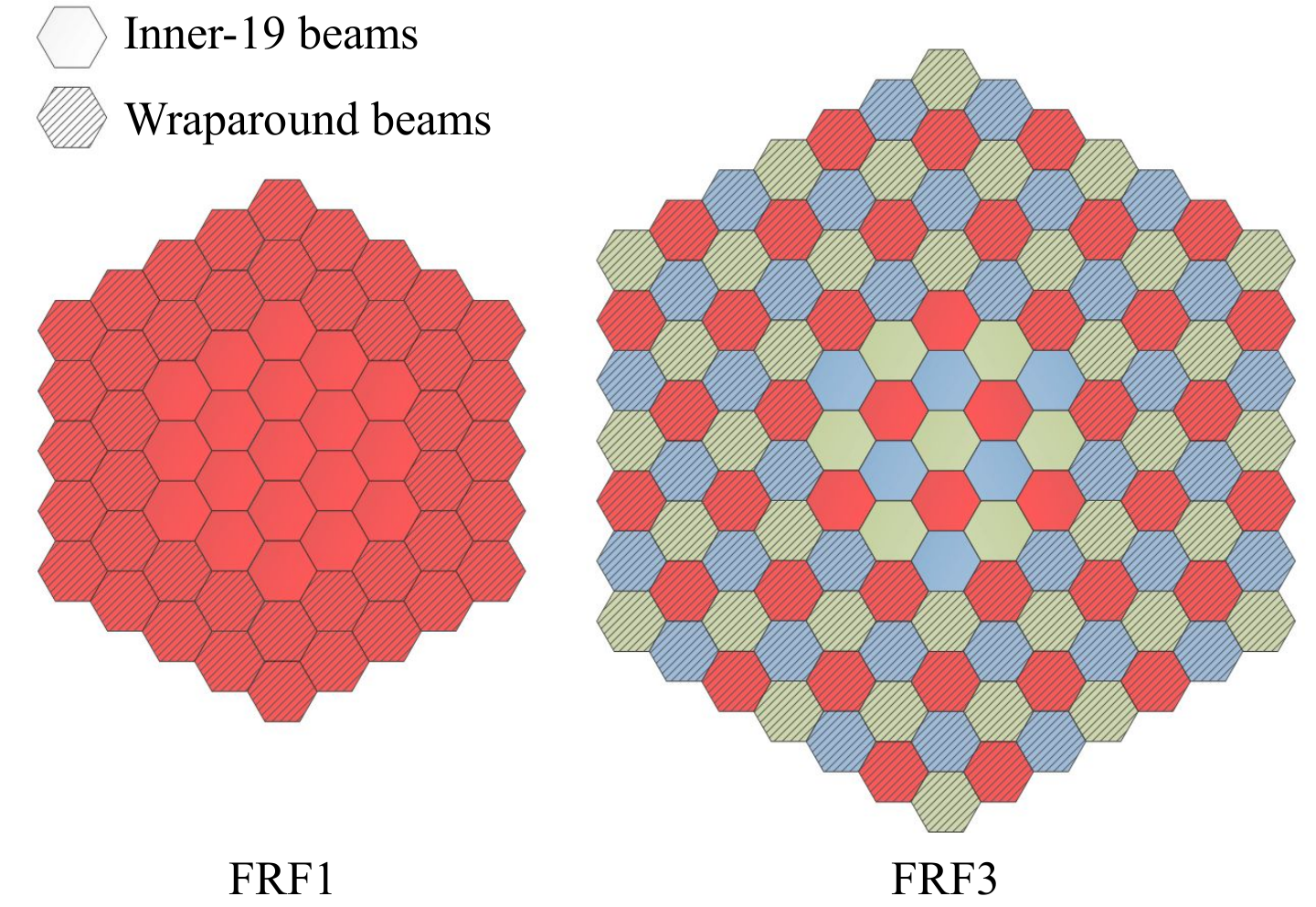}
    \caption{Different frequency reuse schemes illustrated \cite{38821}.}
    \label{fig:fr}
\end{figure}

The system considered consists of a single \gls{leo} satellite with 19 statistics beams, that is, beams from which statistics are collected. Two different \glspl{frf} are considered, namely, FRF1 and FRF3. With FRF1, the frequency is fully reused by the beams. With FRF3, each beam uses only one-third of the total frequency and adjacent beams use different portions of the frequency to minimize inter-cell interference. In addition to the statistics beams, wraparound beams are used to introduce a realistic level of background interference into the system. Two tiers of wraparound beams are used for FRF1 and four tiers for FRF3. FRF3 requires more tiers of wraparound beams because the interfering beams are further away due to the frequency reuse scheme. The different frequency reuse schemes are illustrated in Fig.~\ref{fig:fr}.

Different antenna configurations are considered for users. In the (m,n,p) antenna configuration notation, m and n are the number of \gls{rx} antenna elements in the vertical and horizontal directions, respectively, and p is the number of polarizations used. As described in \hbox{ITU-R M.2514-0}, (1,1,2) and (1,2,2) configurations are considered. The two \gls{rx} antenna polarizations are assumed to revoke the depolarization loss of 3~dB caused by the mismatch between the satellite's \gls{tx} antenna’s single circular polarization against the receiving user equipment's (UE) linearly polarized elements. Additionally, given the different number of UE’s horizontal Rx antenna components (1 or 2), the overall received \gls{sinr} is the \gls{mrc} version of the signal, that is, the average \gls{sinr} of the antenna elements multiplied by the number of them. The signal combination model for multiple Rx antennas is an idealized one, assuming that the neighboring beam interference is noise-like and does not combine at the receiver.

In the \gls{ul} direction, the users have a single \gls{tx} antenna but two different configurations are considered. These are Configuration~A and Configuration~B, as described in Section~6.1.1.1 of TR~38.821. With Configuration~A, similarly to the \gls{dl} direction, a 3~dB depolarization loss is considered. Configuration~A assumes polarization reuse on the satellite side, that is, each beam uses either \gls{rhcp} or \gls{lhcp}. With Configuration~B, the polarization mismatch loss is revoked by the use of two polarizations (\gls{rhcp} and \gls{lhcp}) per beam on the satellite side.

10 UEs are connected to each of the beams, each with full buffer traffic. The UEs considered are handheld devices and are served in \hbox{S-band}.

\begin{table}[]
\caption{Evaluation parameters.}
\label{table:params}
\begin{tabular}{l|l}
{\textbf{\begin{tabular}[c]{@{}l@{}}Technical configuration\\ parameter\end{tabular}}} & { \textbf{Reference value for NR NTN}} \\ \hline
Satellite orbit configuration & LEO, 600 km altitude \\
Beam deployment & Quasi-Earth-Fixed \\
Satellite payload & \begin{tabular}[c]{@{}l@{}}Transparent payload without\\ inter-satellite links \end{tabular}\\
Service link frequency & S-band (2~GHz) \\
Channel bandwidth & 30 MHz \\
3 dB beam width & 4.41° \\
Satellite EIRP density & 34 dBW/MHz \\
Satellite antenna gain & 30 dBi \\
Satellite G/T & 1.1 dB/K \\
Spot beam pattern & \begin{tabular}[c]{@{}l@{}}Hexagonal pattern, 19 statistics\\ beams + 2/4 tiers of interfering\\ beams for FRF1/FRF3\end{tabular} \\
Satellite beam diameter & 50 km \\
Inter-cell distance & 43.3 km \\
TRxP density ($\rho$) & 1/1415 km$^2$ \\
Terminal type & Handheld \\
UE Antenna type & Omni-directional \\
UE antenna polarisation & Linear: ±45°X-pol \\
UE Antenna gain & 0 dBi\\
UE Antenna temperature & 290 K\\
Noise figure & 7 dB \\
UE Tx power & 200 mW (23 dBm) \\
UE Rx antenna configuration & (1,1,2) or (1,2,2) \\
Device deployment & \begin{tabular}[c]{@{}l@{}}100\% outdoor, randomly, and\\ uniformly distributed over the area\end{tabular} \\
UE density & 10 UEs per spot beam \\
UE mobility model & Stationary \\
Traffic model & Full buffer \\
UE antenna height & 1.5 m \\
Satellite antenna pattern & \begin{tabular}[c]{@{}l@{}}Bessel function as in\\ Section 6.4.1 in TR 38.811\end{tabular} \\
\begin{tabular}[c]{@{}l@{}}Satellite antenna\\ polarization configuration\end{tabular} & Circular \\
Central beam elevation & 90 degrees \\
FRF & 1 or 3 \\
Propagation conditions & Line-of-sight probability of 100\% \\
Large-scale channel model & \begin{tabular}[c]{@{}l@{}}Large-scale model of\\ Section 6.6 in TR 38.811\end{tabular} \\
Small-scale channel model & \begin{tabular}[c]{@{}l@{}}Small-scale fading as in\\ Table 6.1.1.1-7  in TR 38.821\end{tabular} \\
Handover margin & 0 dB \\
UE attachment & \begin{tabular}[c]{@{}l@{}}Reference signal received\\ power (RSRP)\end{tabular} \\
Satellite antenna configuration & 1~Rx / 1~Tx per beam \\
Adaptive coding and modulation & Enabled \\
HARQ & Enabled \\
Scheduler & Proportional fair \\
\begin{tabular}[c]{@{}l@{}}Modulation and coding\\ scheme index table\end{tabular} & Table 3 \cite{38214} \\
RNG runs & 5
\\ \hline
\end{tabular}
\end{table}

\subsection{Simulations}
The evaluations are performed using a 5G \gls{ntn} simulator also known as the ALIX simulator \cite{ntn}. The ALIX simulator is a packet-level system simulator. It is based on \gls{ns3} \cite{ns3} and its 5G LENA module \cite{5glena}, which can be used to simulate terrestrial 5G networks. As part of several research projects, multiple components have been developed on top of \gls{ns3} and 5G LENA in the ALIX simulator to enable the simulation of \glspl{ntn}. The \gls{tr}~38.811 channel and beam modeling has been implemented in the simulator and the simulator has been calibrated using the \gls{tr}~38.821 calibration scenarios.

Each different simulation configuration is run five times with unique \gls{rng} seeds, resulting in variations such as differing UE positions. The results for the distribution statistics are combined, while the scalar statistics are averaged. For convenience, the simulation results are summarized in Table~\ref{table:results} where the cases that do not meet the requirements are marked with 'X'. The results capture the effect of using different UE antenna configurations and frequency reuse schemes. In \gls{dl},  the (1,1,2) and (1,2,2) configurations correspond to 1 and 2 UE antennas, respectively.

\begin{figure}[hbt!]
    \centering
    \begin{subfigure}{.49\linewidth}
        \centering
        \includegraphics[width=\linewidth]{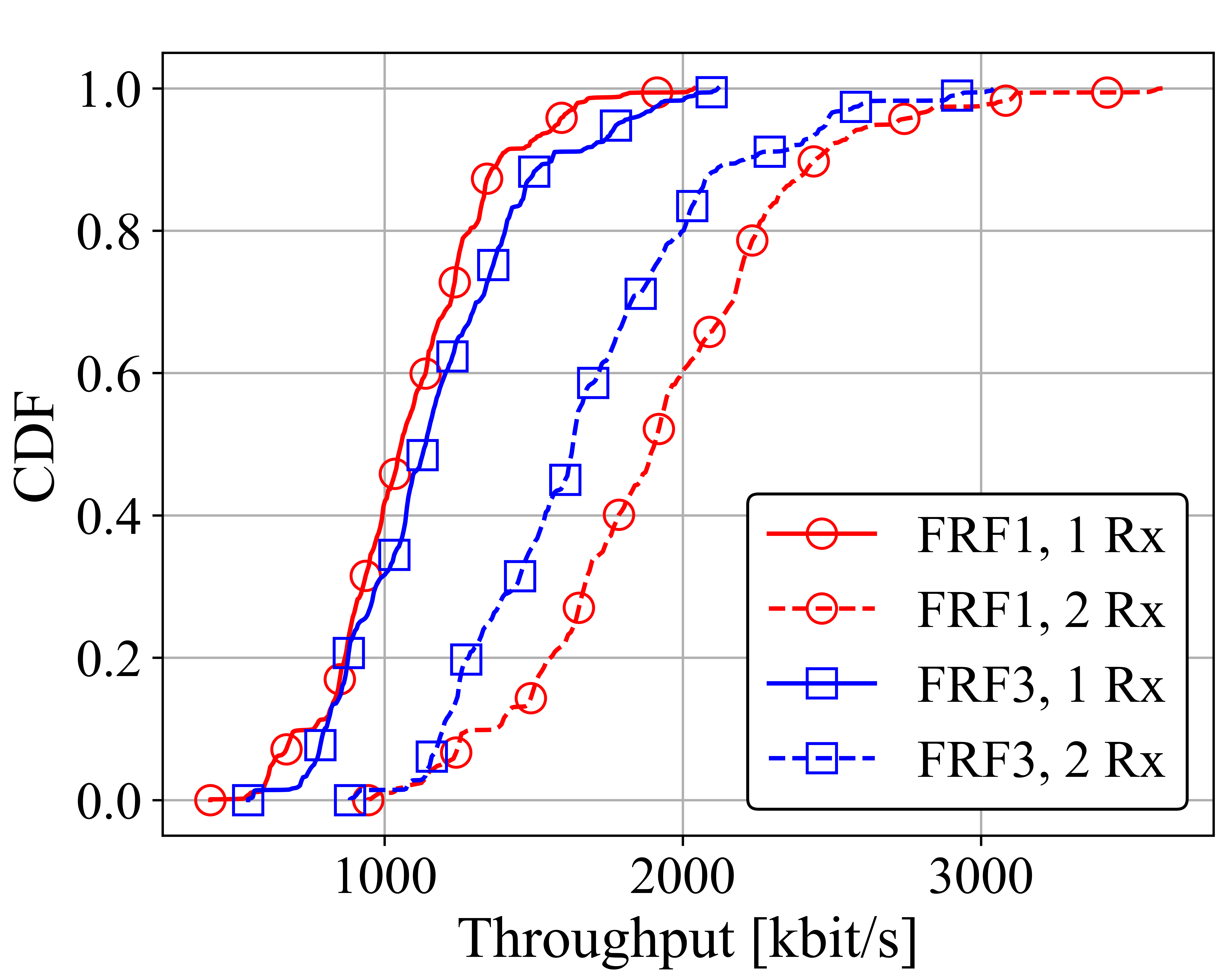}
        \caption{}
        \label{fig:second_sub}
    \end{subfigure}
    \begin{subfigure}{.49\linewidth}
        \centering
        \includegraphics[width=\linewidth]{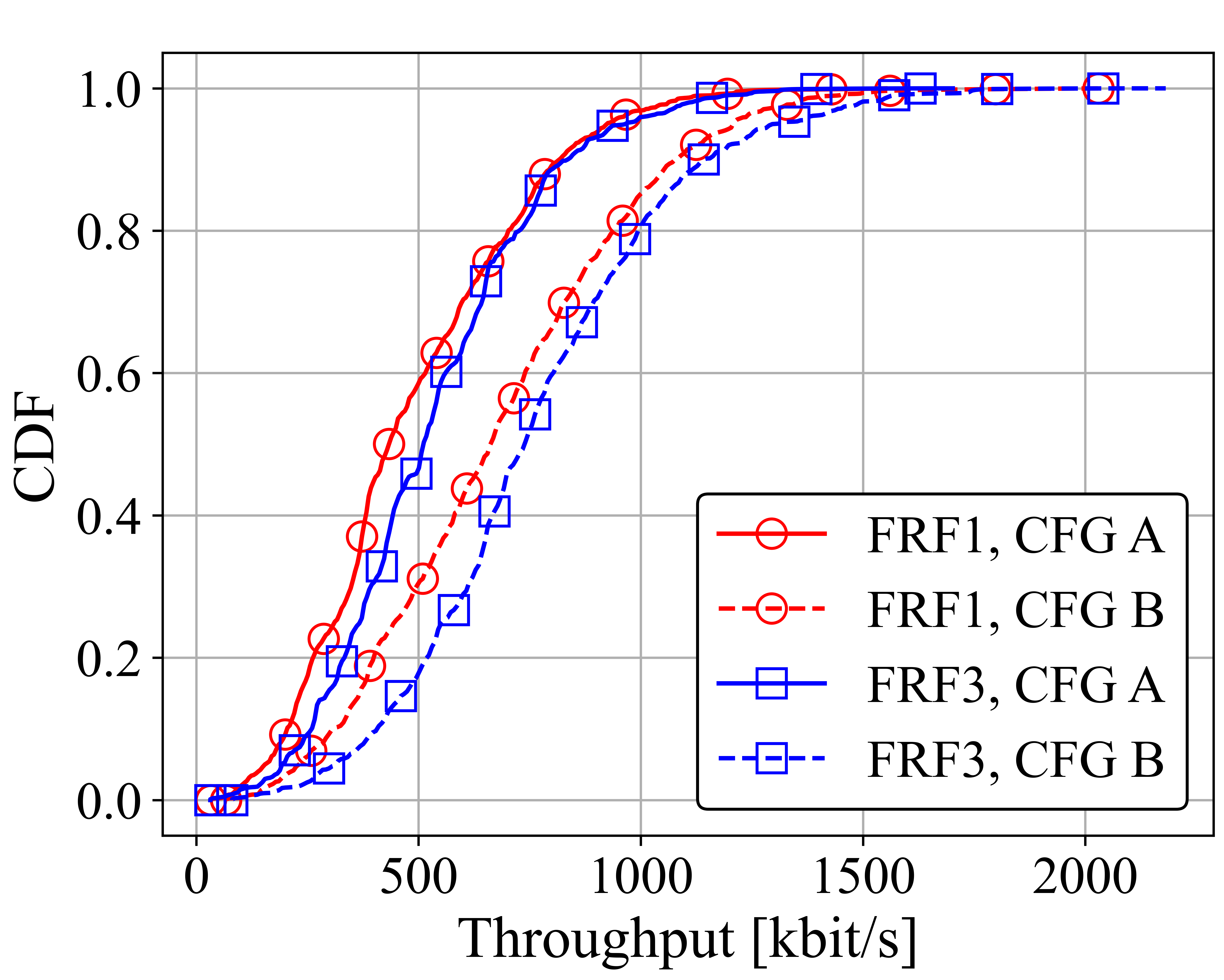}
        \caption{}
        \label{fig:third_sub}
    \end{subfigure}
    \caption{Throughput statistics per UE for a) \gls{dl} and b) \gls{ul}.}
    \label{fig:tp_primary}
\end{figure}

Fig.~\ref{fig:tp_primary}. captures the throughput statistics per UE for a) \gls{dl} and b) \gls{ul}. According to the requirements of \hbox{ITU-R M.2514-0}, the required user-experienced throughput refers to the 5th percentile throughput. This means that the focus is on the lower end of the throughput measurements to ensure that even the users with the poorest performance receive an acceptable service. Meanwhile, the throughput distributions are constructed from the user spectral efficiency statistics by

\begin{equation*}
    R_\textnormal{user} = W \cdot \textnormal{SE}_\textnormal{user},
\end{equation*}

\noindent
where $W$ is the bandwidth and $\textnormal{SE}_\textnormal{user}$ is the UE spectral efficiency.

It is observed that with both frequency reuse schemes, the requirement (1~Mbit/s) cannot be met in \gls{dl} with 1~Rx antenna. However, by increasing the number of UE antennas to 2, the requirements are met. In \gls{ul}, the requirement (100~kbit/s) is met with both frequency reuse schemes and with both Configuration~A and Configuration~B.

\begin{figure}[hbt!]
    \centering
    \begin{subfigure}{.49\linewidth}
        \centering
        \includegraphics[width=\linewidth]{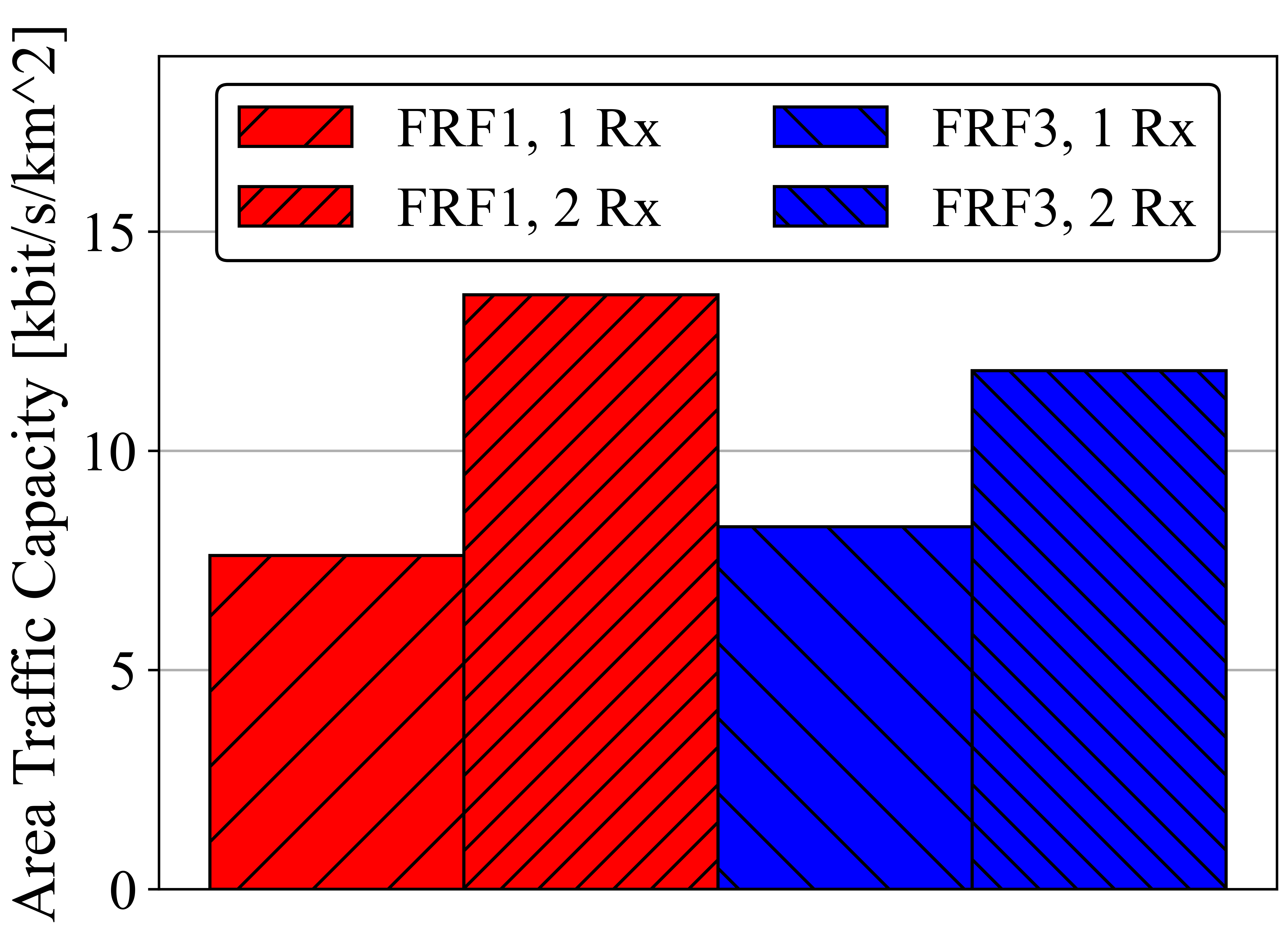}
        \caption{}
        \label{fig:second_sub}
    \end{subfigure}
    \begin{subfigure}{.49\linewidth}
        \centering
        \includegraphics[width=\linewidth]{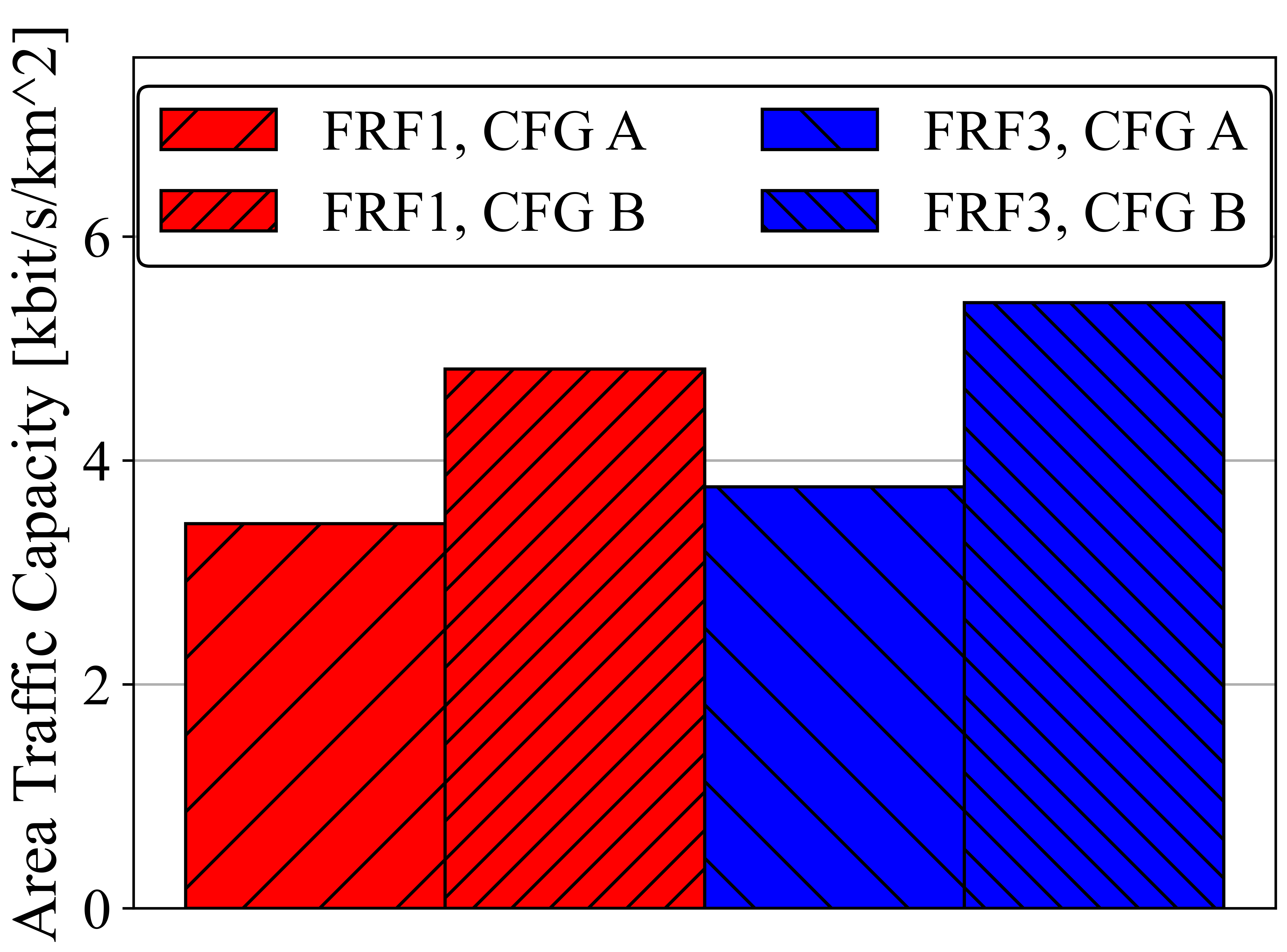}
        \caption{}
        \label{fig:third_sub}
    \end{subfigure}
    \caption{Area traffic capacity statistics for a) \gls{dl} and b) \gls{ul}.}
    \label{fig:atc_primary}
\end{figure}

The area traffic capacity is defined as

\begin{equation*}
    C = \rho \cdot W \cdot \textnormal{SE}_\textnormal{avg},
\end{equation*}

\noindent
where $\rho$ is the transmission point density, $W$ is the system bandwidth, and $\textnormal{SE}_\textnormal{avg}$ is the average spectral efficiency per cell. The area traffic capacity statistics are shown in Fig.~\ref{fig:atc_primary}. for a) \gls{dl} and b) \gls{ul}. The required capacity of~8 kbit/s/km$^2$ is only achieved with FRF3 in the \gls{dl} direction when considering 1~UE Rx antenna. With 2~UE Rx antennas, both frequency reuse schemes meet the requirement. In \gls{ul}, the requirement (1.5~kbit/s/km$^2$) is met by a wide margin, with values more than twice the requirement for Configuration~A and more than three times the requirement for Configuration~B.

\begin{figure}[hbt!]
    \centering
    \begin{subfigure}{.49\linewidth}
        \centering
        \includegraphics[width=\linewidth]{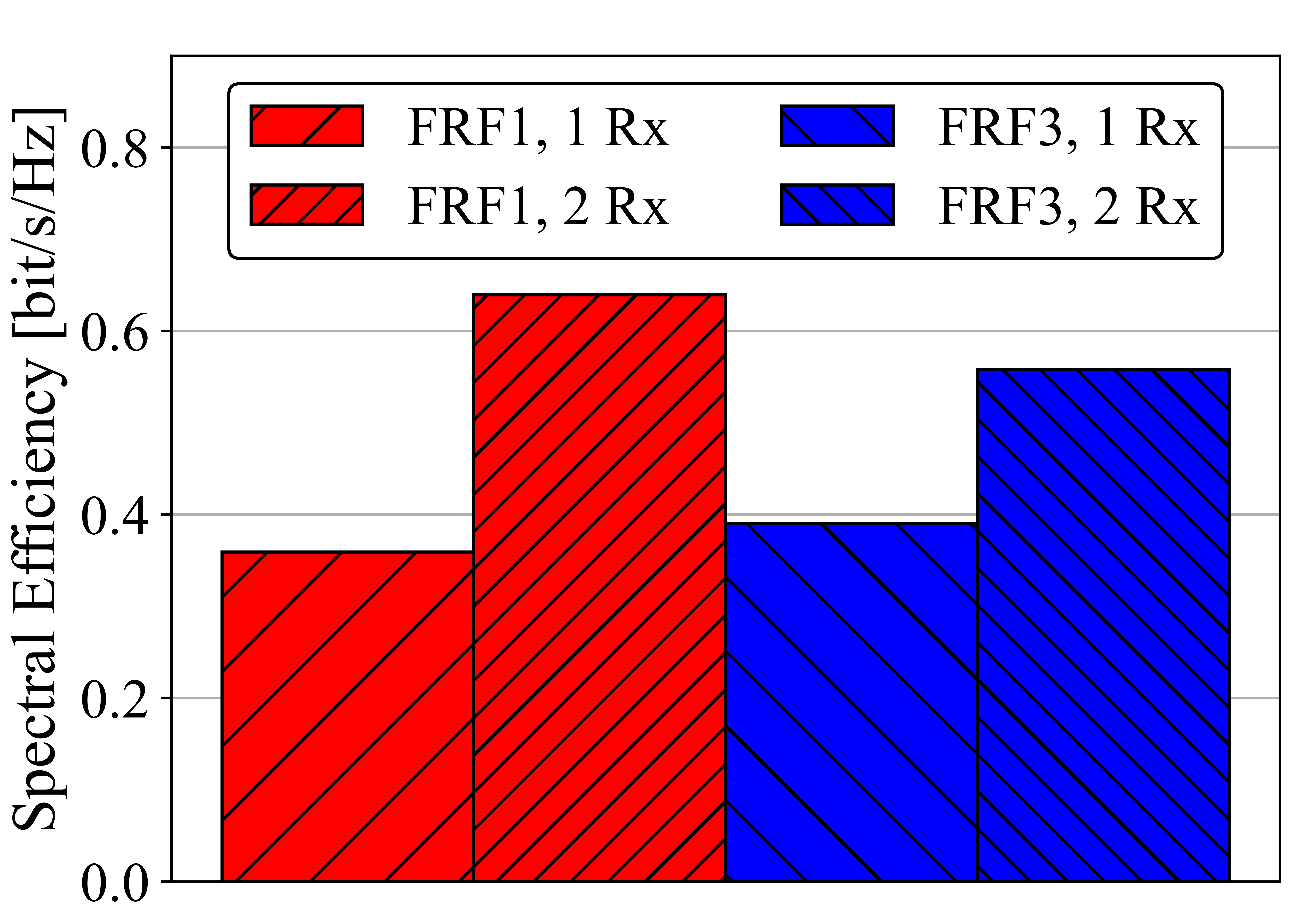}
        \caption{}
        \label{fig:second_sub}
    \end{subfigure}
    \begin{subfigure}{.49\linewidth}
        \centering
        \includegraphics[width=\linewidth]{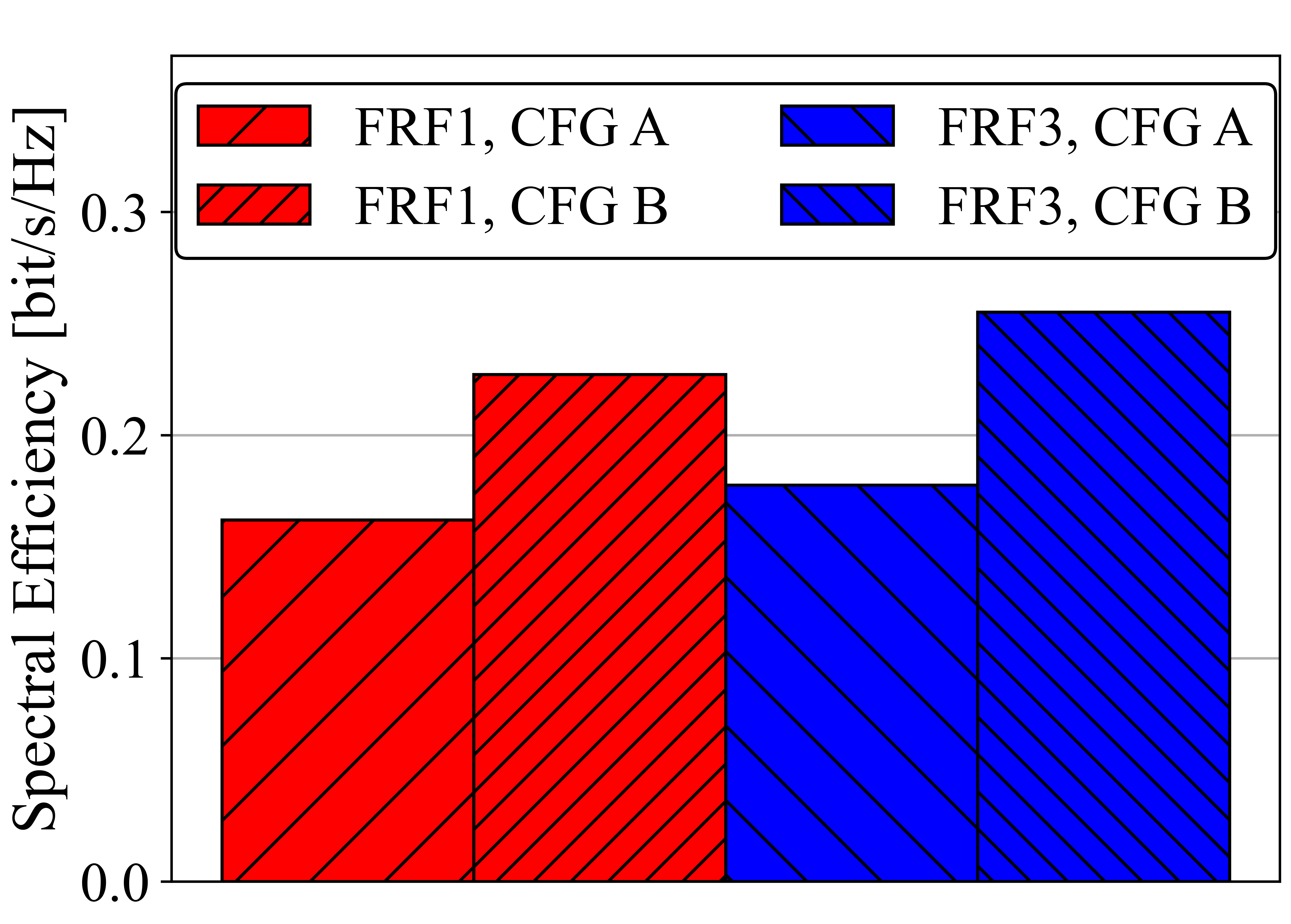}
        \caption{}
        \label{fig:third_sub}
    \end{subfigure}
    \caption{Average spectral efficiency statistics per cell for a) \gls{dl} and b) \gls{ul}.}
    \label{fig:secell_primary}
\end{figure}

The average spectral efficiency statistics per cell are shown in Fig.~\ref{fig:secell_primary}. for a) \gls{dl} and b) \gls{ul}. Again, the \gls{ul} requirement (0.1~b/s/Hz) is far exceeded. In \gls{dl}, 2~UE Rx antennas are required to reach the requirement (0.5~b/s/Hz).

\begin{figure}[hbt!]
    \centering
    \begin{subfigure}{.49\linewidth}
        \centering
        \includegraphics[width=\linewidth]{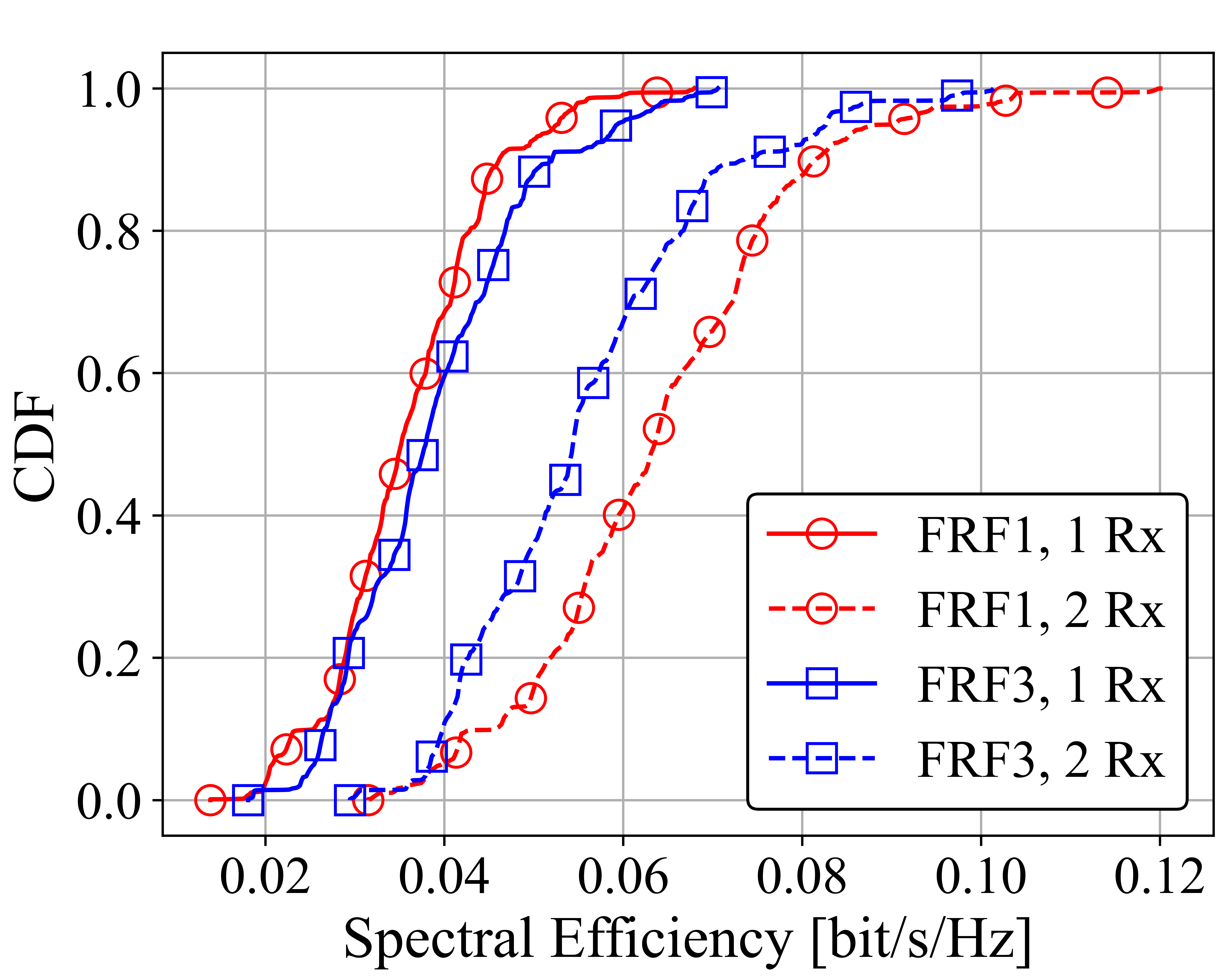}
        \caption{}
        \label{fig:second_sub}
    \end{subfigure}
    \begin{subfigure}{.49\linewidth}
        \centering
        \includegraphics[width=\linewidth]{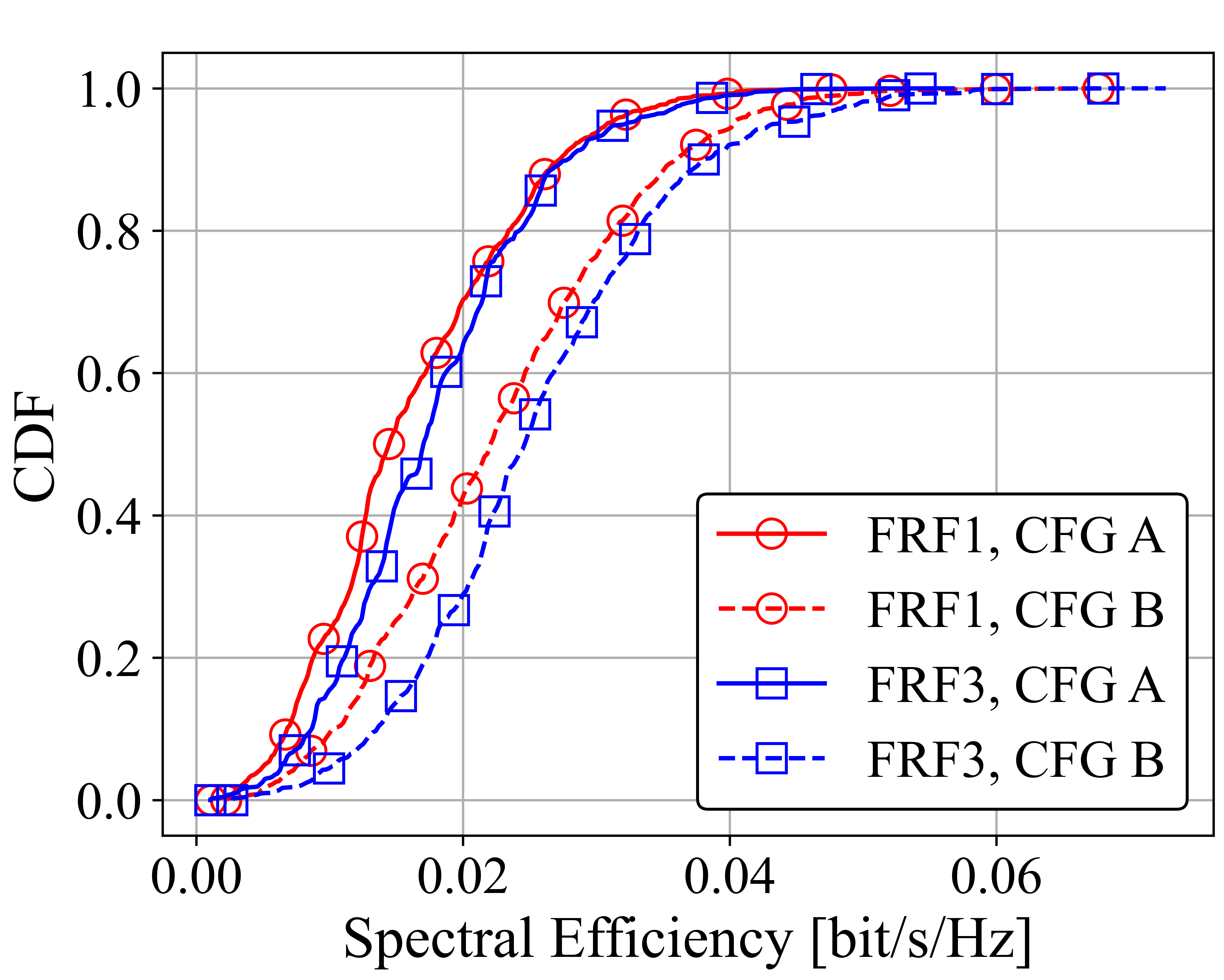}
        \caption{}
        \label{fig:third_sub}
    \end{subfigure}
    \caption{Spectral efficiency statistics per UE for a) \gls{dl} and b) \gls{ul}.}
    \label{fig:seue_primary}
\end{figure}

The spectral efficiency statistics per UE are shown in Fig.~\ref{fig:seue_primary}. for a) \gls{dl} and b) \gls{ul}. The requirement for \gls{dl} is 0.03~b/s/Hz 5th percentile spectral efficiency, while the requirement for \gls{ul} is 0.003~b/s/Hz 5th percentile spectral efficiency. For \gls{dl}, 2~UE Rx antennas are required to meet the requirement regardless of the frequency reuse scheme while the \gls{ul} requirements are met in all considered cases.

\subsection{Effect of Scintillation}

The results presented suggest that 1~Rx UE antenna is not capable of meeting the requirements. However, given the negligible impact of scintillation at mid-latitudes, simulation results with negligible scintillation paired with a single UE Rx antenna are presented below to determine if such a setup could meet the requirements.

Scintillation, which causes rapid signal fluctuations, affects ionospheric propagation mainly below 6~GHz and occasionally up to 10~GHz, while it is most severe at frequencies below 3~GHz. Scintillation varies with location, time, season, and solar activity, being more pronounced post-sunset at low latitudes and at high latitudes. The following simulations consider regions such as the continental United States, Central Europe, and parts of East Asia, which typically fall within mid-latitude ranges where the effect of scintillation is negligible. The effect of scintillation is implemented in the simulator as described in Section~6.6.6 in TR~38.811.

\begin{figure}[hbt!]
    \centering
    \begin{subfigure}{.49\linewidth}
        \centering
        \includegraphics[width=\linewidth]{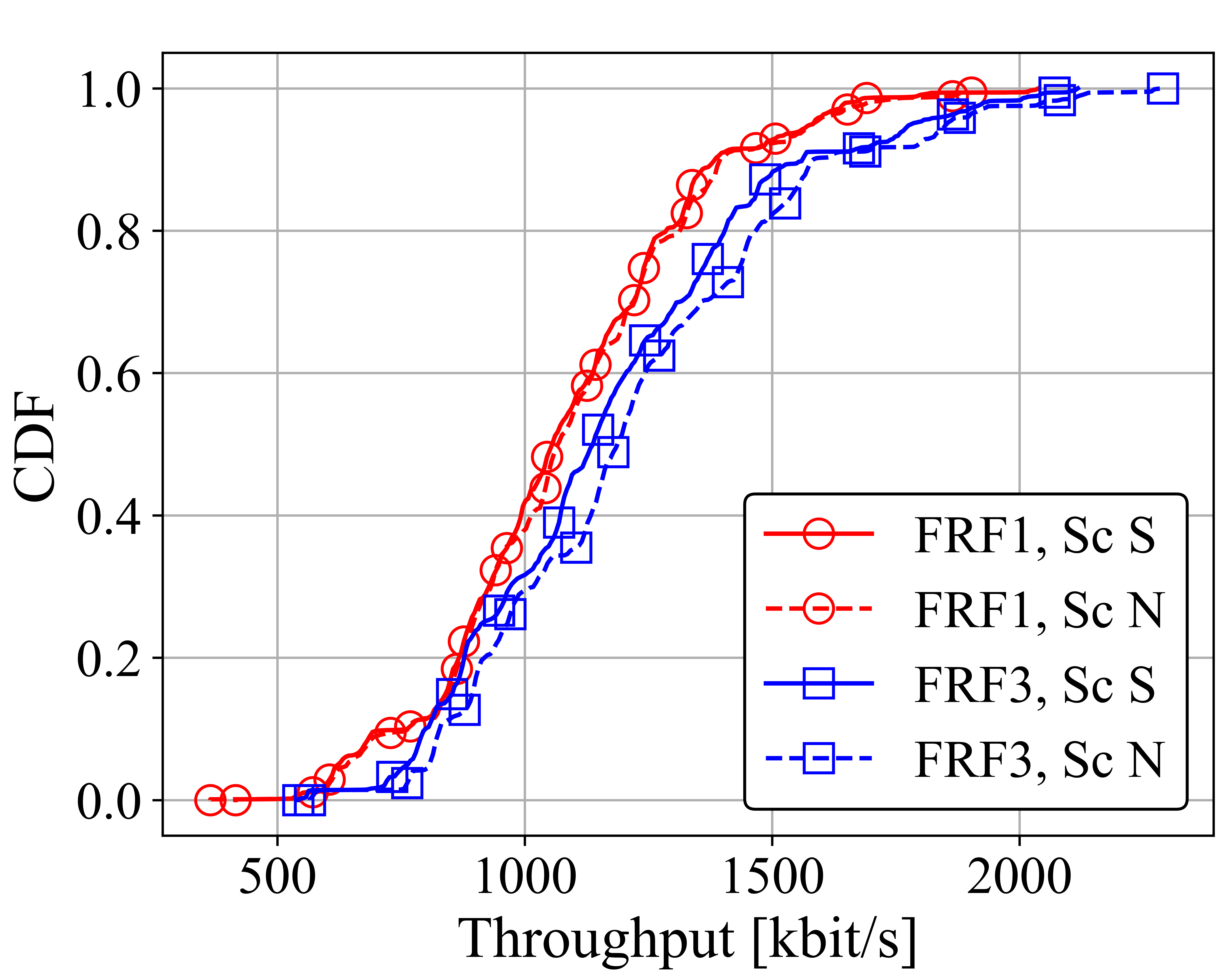}
        \caption{}
        \label{fig:second_sub}
    \end{subfigure}
    \begin{subfigure}{.49\linewidth}
        \centering
        \includegraphics[width=\linewidth]{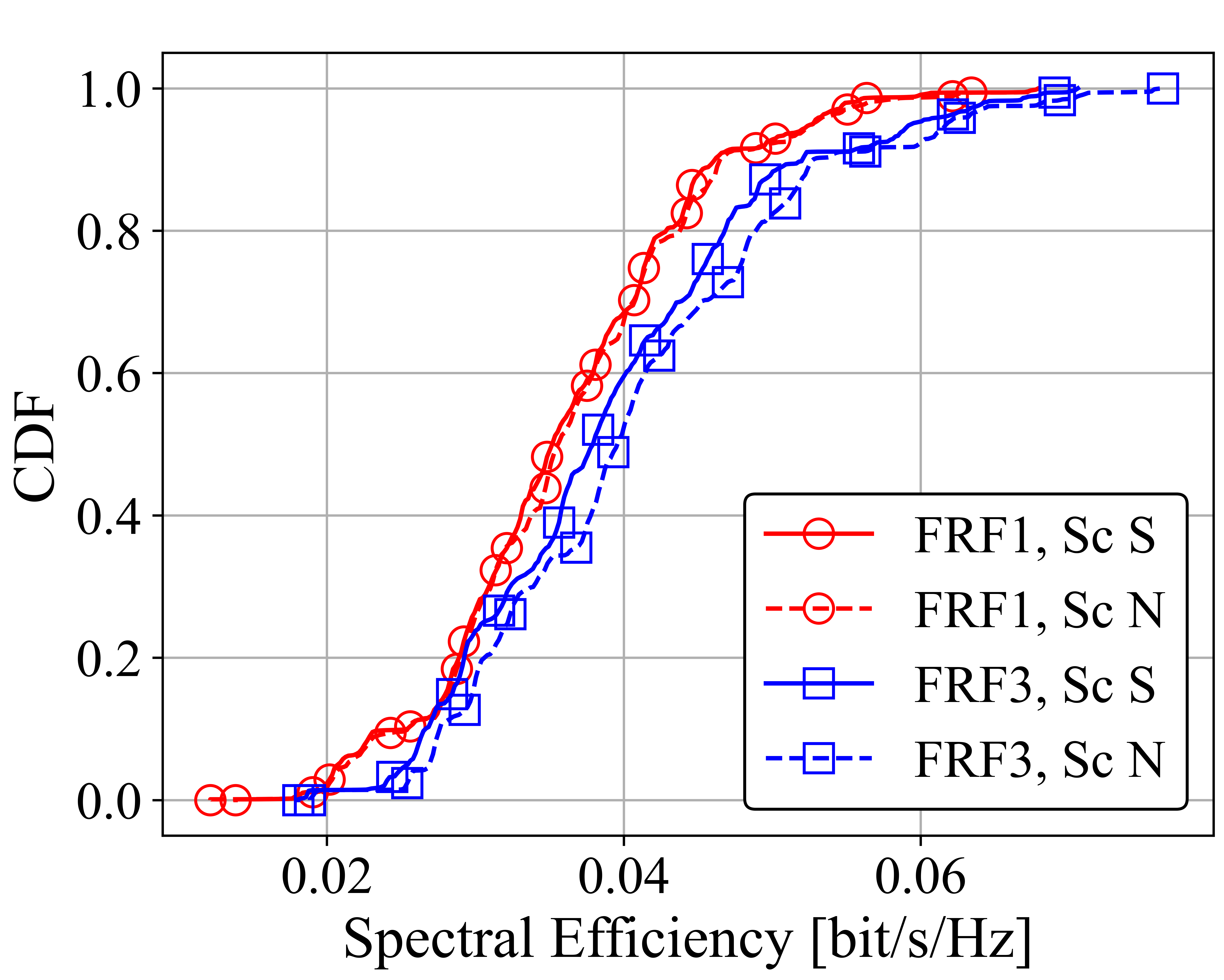}
        \caption{}
        \label{fig:third_sub}
    \end{subfigure}
    \caption{a) Throughput and b) spectral efficiency statistics per UE for \gls{dl} with 1~Rx antenna showing the effect of scintillation.}
    \label{fig:tp_se_secondary}
\end{figure}

Fig.~\ref{fig:tp_se_secondary} illustrates how scintillation affects throughput and spectral efficiency per UE. In the figure, the "Sc~S" label corresponds to the significant scintillation case, and the "Sc~N" label corresponds to the negligible scintillation case. The figure reveals that for FRF1, scintillation has little effect. For FRF3, there is a slight improvement in performance when scintillation is negligible. This is because, with FRF1, UEs experience interference as the primary limiting factor rather than noise. Scintillation, which acts as an additional fading component, affects both desired and interfering signals. Consequently, reducing scintillation does not significantly change the SINR because both signal and interference are affected proportionally by the scintillation loss, keeping the SINR at approximately the same level.

\begin{figure}[hbt!]
    \centering
    \begin{subfigure}{.49\linewidth}
        \centering
        \includegraphics[width=\linewidth]{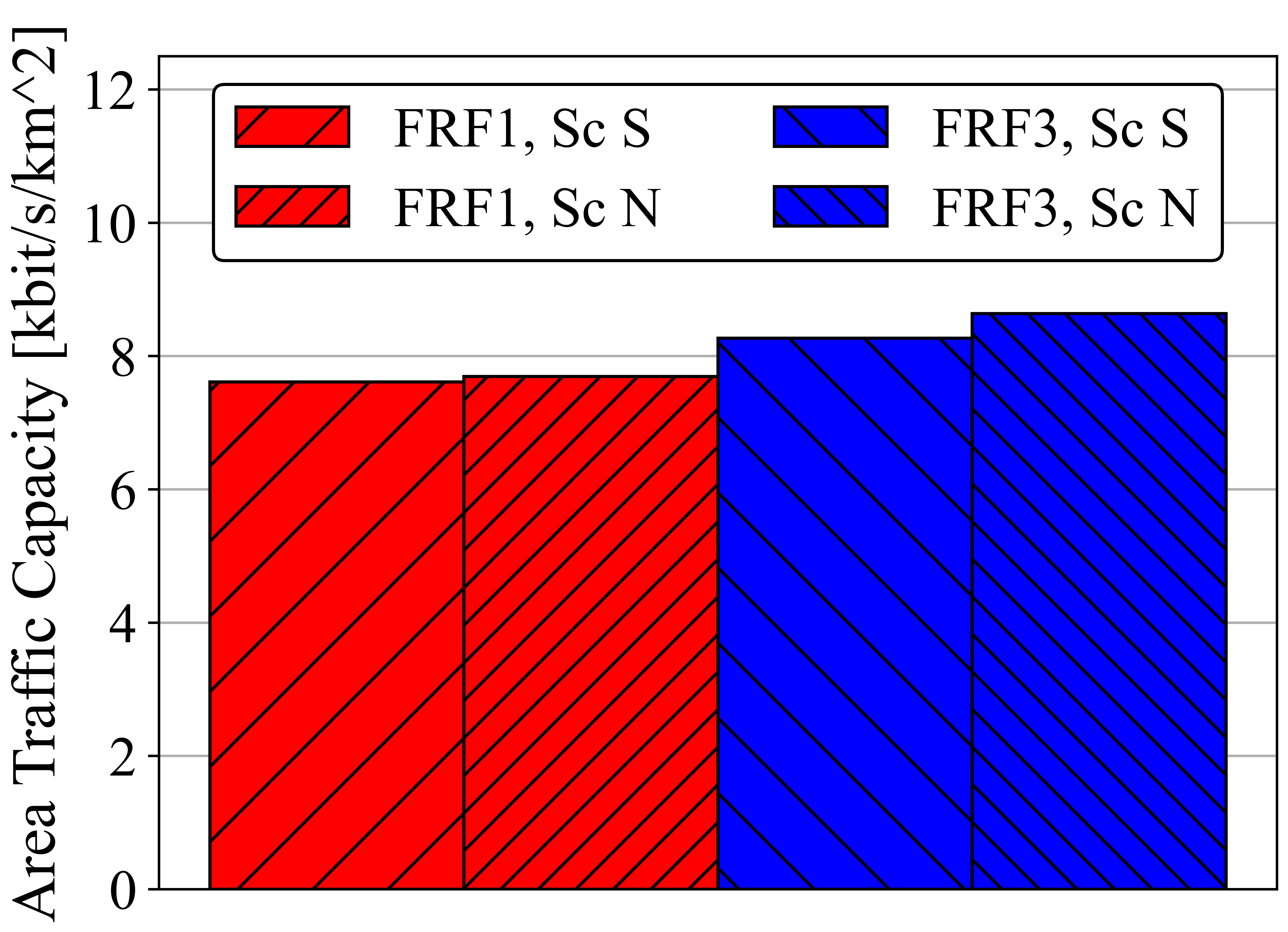}
        \caption{}
        \label{fig:second_sub}
    \end{subfigure}
    \begin{subfigure}{.49\linewidth}
        \centering
        \includegraphics[width=\linewidth]{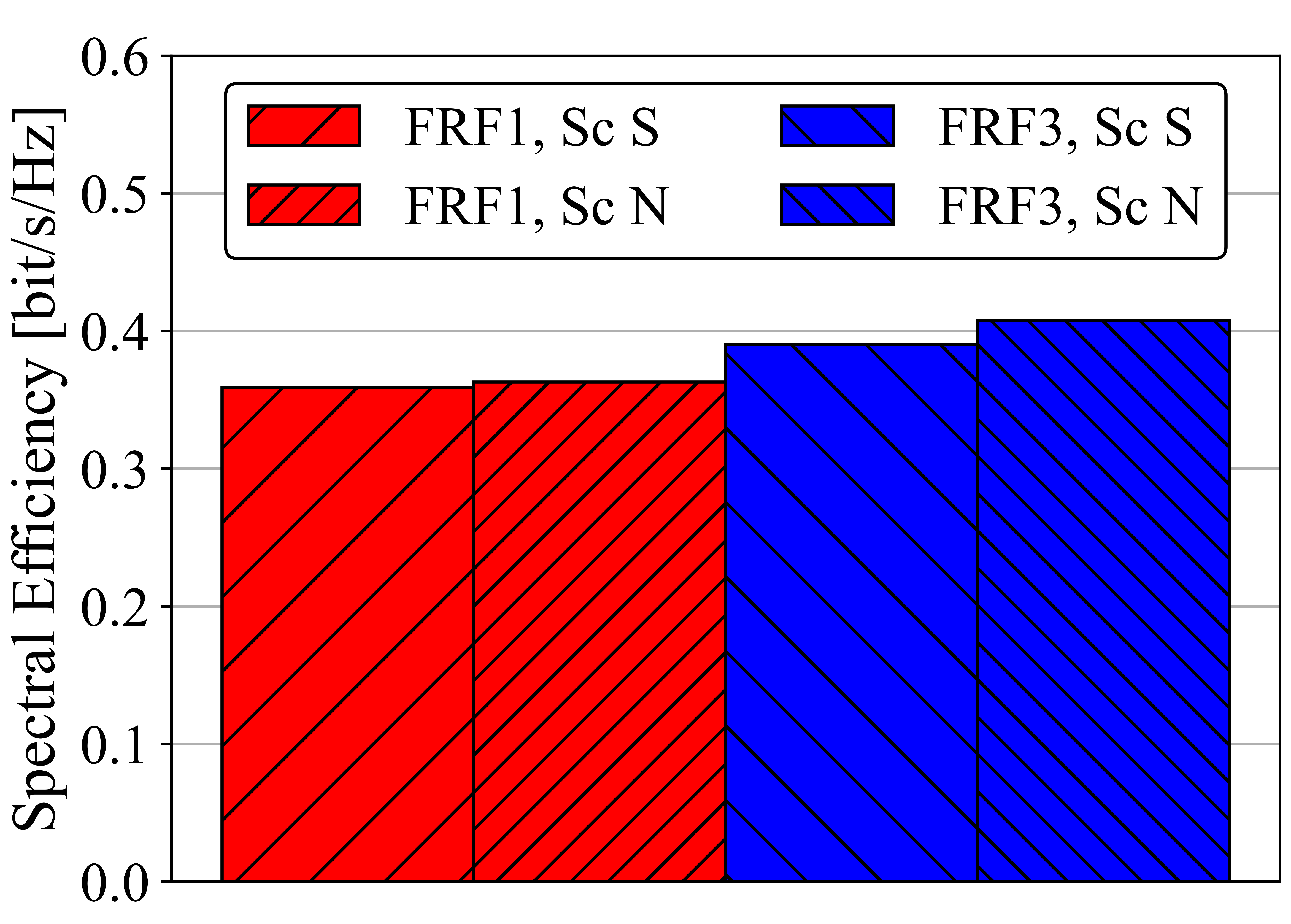}
        \caption{}
        \label{fig:third_sub}
    \end{subfigure}
    \caption{a) Area traffic capacity and b) average spectral efficiency statistics per cell for \gls{dl} with 1~Rx antenna showing the effect of scintillation.}
    \label{fig:atc_se_secondary}
\end{figure}

Conversely, for FRF3, where noise is more dominant than interference, the SINR is more sensitive to changes in scintillation. In these cases, the SINR is greater with negligible scintillation than the SINR with significant scintillation. Therefore, removing the fading effect of scintillation can result in a small performance improvement in such noise-sensitive scenarios. Fig.~\ref{fig:atc_se_secondary} supports this observation, showing that the area traffic capacity and the average spectral efficiency per cell follow the same pattern when scintillation effects are considered.

\begin{table*}[hbt!]
\begin{center}
\caption{Simulation results. The cases that do not meet the requirements are marked with 'X'.}
\label{table:results}
\begin{tabular}{|l|l|l|l|l|l|l|l|l|}
\hline
\textbf{Link direction} & \textbf{\begin{tabular}[c]{@{}l@{}}Number\\ of UE   \\ antennas\end{tabular}} & \textbf{FRF} & \textbf{Scintillation} &  & \textbf{\begin{tabular}[c]{@{}l@{}}User experienced\\ data rate {[}Mbit/s{]}\end{tabular}} & \textbf{\begin{tabular}[c]{@{}l@{}}5th   percentile\\ spectral efficiency\\ {[}bit/s/Hz{]}\end{tabular}} & \textbf{\begin{tabular}[c]{@{}l@{}}Average spectral\\ efficiency\\ {[}bit/s/Hz{]}\end{tabular}} & \textbf{\begin{tabular}[c]{@{}l@{}}Area traffic\\ capacity\\ {[}kbit/s/km$^2${]}\end{tabular}} \\ \hline
\multirow{7}{*}{\gls{dl}} & \multirow{2}{*}{1} & 1 & \multirow{4}{*}{Significant} &  & 0.62 \xmark & 0.021 \xmark & 0.36 \xmark & 7.6 \xmark \\ \cline{3-3} \cline{5-9} 
 &  & 3 &  &  & 0.77 \xmark& 0.026 \xmark & 0.39 \xmark & 8.3 \\ \cline{2-3} \cline{5-9} 
 & \multirow{2}{*}{2} & 1 &  &  & 1.20 & 0.040 & 0.64 & 13.6 \\ \cline{3-3} \cline{5-9} 
 &  & 3 &  &  & 1.15 & 0.038 & 0.56 & 11.8 \\ \cline{2-9} 
 & \multirow{2}{*}{1} & 1 & \multirow{2}{*}{Negligible} &  & 0.64 \xmark & 0.021 \xmark & 0.36 \xmark & 7.7 \xmark \\ \cline{3-3} \cline{5-9} 
 &  & 3 &  &  & 0.81 \xmark & 0.027 \xmark & 0.41 \xmark & 8.6 \\ \cline{2-9} 
 &  &  &  & Required & 1.0 & 0.03 & 0.5 & 8.0 \\ \hline
\multirow{2}{*}{\gls{ul},   CFG A} & \multirow{2}{*}{1} & 1 & \multirow{4}{*}{Significant} &  & 0.16 & 0.0052 & 0.16 & 3.4 \\ \cline{3-3} \cline{5-9} 
 &  & 3 &  &  & 0.19 & 0.0064 & 0.18 & 3.8 \\ \cline{1-3} \cline{5-9} 
\multirow{2}{*}{\gls{ul},   CFG B} & \multirow{2}{*}{1} & 1 &  &  & 0.23 & 0.0077 & 0.23 & 4.8 \\ \cline{3-3} \cline{5-9} 
 &  & 3 &  &  & 0.32 & 0.011 & 0.26 & 5.4 \\ \hline
\gls{ul} &  &  &  & Required & 0.1 & 0.003 & 0.1 & 1.5 \\ \hline
\end{tabular}
\end{center}
\end{table*}

\subsection{Result Summary}

The results indicate that with a single Rx antenna, only the area traffic capacity requirement is achievable, both with significant and negligible scintillation and with FRF3. Even in scenarios with negligible scintillation, representative of mid-latitude regions such as Central Europe, the requirements are not met with a single Rx antenna. This illustrates the potential challenge of meeting the requirements with a single Rx antenna.

When the number of Rx antennas is increased to two, all the considered requirements are met with both FRF1 and FRF3, with FRF1 showing higher overall performance than FRF3, which is a surprising finding, as FRF3 is typically considered to perform better than FRF1 due to better minimization of interference. This may be explained by the product of the improved spectral efficiency of the dual-antenna modulation and coding scheme and the full bandwidth for FRF1 being greater than the equivalent product for FRF3.

The implementation of additional Rx antennas may play a critical role in meeting the performance requirements. However, the combination gain in this study is an idealized version assuming interference from neighboring beams does not combine constructively at the receiver. The actual gain of the signal combination scheme is left for further study, although it is safe to assume that the use of antenna diversity improves the overall channel quality.

The \gls{ul} requirements are consistently met even with Configuration~A, which is a more challenging setup. This is particularly promising because it indicates that the requirements can be met even with polarization reuse. \gls{ul} communications, which are generally more demanding, benefit from this observation. Due to the consistent compliance with \gls{ul} requirements in all cases considered, results for \gls{ul} with negligible scintillation are not detailed in the study. However, it is inferred that in such setups, \gls{ul} could exceed the requirements by an even greater margin, further demonstrating the robustness of the system under varying conditions.

\section{Conclusion}
\label{sec:conclusion}

In this paper, the \gls{imt2020} requirements for satellite radio interface technology were evaluated with a focus on the requirements for throughput, area traffic capacity, and spectral efficiency statistics in non-mobile \hbox{eMBB-s} scenarios. It was observed that the \gls{dl} requirements can be met with 2~Rx antennas but not with 1~Rx antenna even when scintillation is negligible while the \gls{ul} requirements can be met with a large margin with 1~Tx antenna with both Configuration~A (beam polarization reuse) and B (no beam polarization reuse). Furthermore, the results indicate that FRF1 may outperform FRF3 in DL with a dual-antenna setup, which is a surprising finding since FRF3 is typically considered to outperform FRF1 due to better interference reduction.

A possible future work could be the analysis of the compliance of 3GPP NR NTNs with the other \gls{imt2020} requirements for satellite radio interface technology. In addition, it would be important to investigate whether the requirements could somehow be met with 1~Rx antenna, for example, by scheduling optimization, advanced antenna designs, new modulation schemes, or AI-driven network optimization. Furthermore, it was assumed that interference from neighboring beams does not constructively combine at the receiver, a topic that warrants further investigation in future work.

In conclusion, the research indicates that 3GPP NR NTNs comply with the ITU's requirements, supporting further standardization and deployment efforts. This development paves the way for seamless integration of terrestrial and non-terrestrial networks, aiding the transition to 6G and enabling global connectivity. It promises improved access in remote areas and supports multiple applications in broadband, emergency services, and IoT. However, it also highlights challenges in technology harmonization and equitable spectrum allocation, underscoring the need for collaborative solutions.

\vspace{6pt}
\bibliography{references} 
\bibliographystyle{IEEEtran}

\end{document}